\documentclass[journal,twoside]{IEEEtran}
\usepackage[dvipsnames]{xcolor}
\usepackage[inline]{enumitem}
\usepackage{amssymb}
\usepackage{dsfont}
\usepackage{subfigure}
\usepackage{bbm}
\usepackage{amsmath}
\usepackage{upgreek}
\usepackage{setspace}
\usepackage{pgfplots}
\usepackage{psfrag}
\usepackage{tikz}
\usepackage{circuitikz}
\usepackage{mathrsfs}
\usepackage{amsfonts}
\usepackage{graphicx,cite,multirow,rotating}
\usepackage{xfrac}
\usepackage{hyperref,stmaryrd}
\usepackage{amsmath}
\usepackage{textcomp}
\usepackage{pgf}
\usepackage{cuted} 
\usepackage{algorithm}
\usepackage{algorithmic}
\usepackage{url}
\usepackage{amsthm} 
\usepackage{thmtools}
\usepackage{caption}
\usepackage{comment}

\def\d={{\stackrel {\rm def}{=}}}
\def\bmu{{\boldsymbol{\mu}}}

\newcommand{\beq}{\begin{equation}}

\newcommand{\eeq}{\end{equation}}
\newcommand{\bqy}{\begin{eqnarray}}
\newcommand{\eqy}{\end{eqnarray}}
\newcommand{\bqi}{\begin{IEEEeqnarray}}
\newcommand{\eqi}{\end{IEEEeqnarray}}
\newcommand{\bb}[1]{{\mathbf #1}}

\newcommand{\XX}[1]{{\mathbf{X}}}

\newcommand{\YY}[1]{{\mathbf{Y}}}
\newcommand{\RR}[1]{{\mathbf{R}}}

\def\balpha{\boldsymbol{\alpha}}
\def\BD{\begin{displaymath}}
\def\BEA{\begin{eqnarray}}
\def\BEAs{\begin{eqnarray*}}
\def\ED{\end{displaymath}}
\def\EE{\end{equation}}
\def\EEA{\end{eqnarray}}
\def\EEAs{\end{eqnarray*}}

\def\bA{\textbf{A}}

\def\bA{{\bf A}}

\def\bB{{\bf B}}
\def\bb{{\bf b}}

\def\bI{{\bf I}}

\def\bM{{\bf M}}

\def\bS{{\bf S}}

\def\bT{{\bf T}}
\def\bt{{\bf t}}

\def\bx{{\bf x}}

\def\by{{\bf y}}
\def\bZ{{\bf Z}}
\def\bz{{\bf z}}

\def\bA{\textbf{A}}

\def\bB{\textbf{B}}
\def\bb{\textbf{b}}

\def\bI{\textbf{I}}

\def\bM{\textbf{M}}

\def\bS{\textbf{S}}

\def\bT{\textbf{T}}
\def\bt{\textbf{t}}

\def\bx{\textbf{x}}

\def\by{\textbf{y}}
\def\bZ{\textbf{Z}}
\def\bz{\textbf{z}}
\def\b0{\textbf{0}}
\def\b1{\textbf{1}}

\def\cH{\mathcal{H}}
\DeclareMathOperator{\Tr}{Tr}

\def\d={{\stackrel{\rm def}{=}}}

\def\W-Rao{{\rm W-Rao}}
\def\NS-Rao{{\rm NS-Rao}}

\def\R{{\mathds R}}

\def\C{{\mathds C}}

\def\balpha{\mbox{\boldmath $\alpha$}}

\def\b_eta{\mbox{\boldmath $\eta$}}

\def\bnu{\mbox{\boldmat $\nu$}}

\def\cH{{\cal H}}

\def\cL{{\cal L}}

\def\L2{L^{2}[-{\pi \over 2} , {\pi \over 2}]}

\def\bnu{\bnu}

\def\R{{\mathds R}}
\newcommand{\tr}{\mbox{\rm Tr}\, }

\newcommand{\ee}{\end{equation}}
\newcommand{\ds}{\displaystyle}
\newcommand{\test}{\mbox{$
\begin{array}{c}
\stackrel{ \stackrel{\textstyle H_{1}}{\textstyle >} }{
\stackrel{\textstyle <}{\textstyle H_0} }
\end{array}
$}}
\newcommand{\dmax}{\begin{displaystyle}\max\end{displaystyle}}

\DeclareMathAlphabet{\mathpzc}{OT1}{pzc}{m}{it}
\title{ {Multiple Sub-Pixel Target Detection for Hyperspectral Imaging Systems}}

\author{Pia Addabbo, \IEEEmembership{Senior Member, IEEE}, Nicomino Fiscante, \IEEEmembership{Student Member, IEEE},\\ Gaetano Giunta, \IEEEmembership{Senior Member, IEEE}, 
Danilo Orlando, \IEEEmembership{Senior Member, IEEE},\\ Giuseppe Ricci, \IEEEmembership{Senior Member, IEEE}, and Silvia Liberata Ullo, \IEEEmembership{Senior Member, IEEE}

\thanks{\emph{(Corresponding author: Pia Addabbo.)}}
\thanks{Pia Addabbo is with Universit\`a degli Studi ``Giustino Fortunato'', 
viale Raffale Delcogliano, 12, 82100 Benevento, Italy
E-mail: {\tt p.addabbo@unifortunato.eu}.}
\thanks{Nicomino Fiscante and Gaetano Giunta are with Industrial, Electronic and Mechanical Engineering Department, 
University of Roma Tre, via Vito Volterra 62, 00146 Rome, Italy. 
E-mail: {\tt nicomino.fiscante@uniroma3.it, gaetano.giunta@uniroma3.it}.}
\thanks{Danilo Orlando is with Universit\`a degli Studi ``Niccol\`o Cusano'',
via Don Carlo Gnocchi 3, 00166 Roma, Italy. 
E-mail: {\tt danilo.orlando@unicusano.it}.}
\thanks{Giuseppe Ricci is with the Dipartimento di Ingegneria dell'Innovazione,
        Universit\`a del Salento, Via Monteroni, 73100 Lecce, Italy.
        E-mail: {\tt giuseppe.ricci@unisalento.it}.}
\thanks{Silvia Liberata Ullo is with Universit\`a degli Studi del Sannio, 
piazza Roma, 21, 82100 Benevento, Italy
E-mail: {\tt ullo@unisannio.it}.}
}

\begin{document}

\maketitle

\begin{abstract}
Hyperspectral target detection is a task of primary importance in remote sensing since it allows 
identification, location, and discrimination of target features. To this end, the reflectance maps, which
contain the spectral signatures and related abundances of the materials in the observed scene, are often used.
However, due to the low spatial resolution of most hyperspectral sensors, targets occupy a fraction of the pixel and, hence,
the spectra of different sub-pixel targets (including the background spectrum) are mixed together within the same pixel.
To solve this issue, in this paper, we adopt a generalized replacement model accounting for 
multiple sub-pixel target spectra and formulate the detection problem at hand as 
a binary hypothesis test where under the alternative hypothesis
the target is modeled in terms of a linear combination of endmembers whose coefficients also account
for the presence of the background. Then, we devise detection architectures
based upon the generalized likelihood ratio test where the unknown parameters are suitably estimated
through procedures inspired by the maximum likelihood approach.
The performances of the proposed decision schemes are evaluated by means of both synthetic as well as real data and compared with an analogous counterpart by showing the effectiveness of the proposed procedure. 
\end{abstract}

\section{Introduction}\label{Section_Intro}
Hyperspectral imaging spectrometers enable identification and discrimination of different target features in a scene
due to hundreds or thousands of spectral channels covering the visible, near and shortwave infrared and ultraviolet spectral bands.
Their field of application is very wide and ranges from agricultural remote sensing, object classification, 
atmospheric monitoring, to military investigation \cite{9361697,9328197,9674932,8738016}. 

On the other hand, the consequent low spatial resolution entails a challenging situation due to the fact that different materials can jointly occupy a single pixel under test (PUT). 
As a matter of fact, the spectra of
different sub-pixel targets (including the background spectrum)
are mixed together as well as the corresponding fraction (or abundance) of constituent endmembers.
In general, the number of endmembers and their abundances at each pixel are unknowns 
and the corresponding estimation process, i.e., the hyperspectral unmixing, gets complicated 
due to the model inaccuracies, the observation noise, the environmental conditions, 
and the endmember variability \cite{974727}. 

Unmixing algorithms currently rely on mixing models that can be either linear or nonlinear. 
The first case corresponds to a macroscopic mixing scale, whereas the second one is more representative of the physical interactions between the scattering from multiple materials. 
As for the unmixing methods, signal-subspace, geometrical, statistical, sparsity-based, and 
spatial-contextual procedures have been proposed over the years \cite{6200362}. 

Recent advances, in the field of hyperspectral imaging, are directed towards the development 
of target detection algorithms fed by hyperspectral images and  exploiting
spectral signatures of the materials to identify the targets of interest \cite{6678247}. 
In this case, the separation of the background signature from the desired targets represents 
the major challenge and the actual classification procedures are not directly 
applicable to target detection since the targets' number is typically too small for using 
clustering-based algorithms. Moreover, the targets of interest may appear as sub-pixel 
targets where the background interference directly distorts the shape of the real observed target spectrum.

With reference to this latter issue, different solutions have been proposed for target detection 
in hyperspectral imaging \cite{6678280}. The main difference between the various algorithms relies on
the availability of prior knowledge about the spectral characteristics of the desired targets.
When the target spectral information is not \textit{a-priori} known, or is affected by 
uncertainty, anomaly detectors can be used, where hyperspectral image anomalies 
are related to a general kind of spectral irregularity due to the presence of atypical objects. 
In this case, pattern recognition or statistical schemes are used for the detection of the objects 
that stand out from the background \cite{9532003}. 
On the contrary, if the spectral characteristics of the desired targets are \textit{a-priori} known, 
both the noise and the background can be statistically modeled as Gaussian-distributed 
and several classical target detection algorithms  can be used, such as the linear 
spectral matched filter (SMF), the matched subspace detector (MSD), 
the adaptive subspace detector (ASD), and the orthogonal sub-space projection (OSP) \cite{974724}.
However, these detectors do not consider any constraint on the abundance of sub-pixel targets and 
background. Otherwise stated, they do account for the fact that when a sub-pixel target is present, 
the amount of background should be reduced by the same proportion, which leads to the definition 
of the so called {\em replacement model}, by which a sub-pixel 
target is supposed to ``replace" or fill part of the background within a given pixel
\cite{rep_model}. It is important to notice that this problem is not a classical detection one, 
as the background power is different under the two hypotheses (background-only versus target-plus-background).
Recent efforts for the development of detectors based on the replacement model can 
be found in \cite{8964590,VINCENT2021108212}.
In \cite{8964590}, the analogous of Kelly's Generalized Likelihood Ratio Test (GLRT) \cite{kelly1986adaptive}
for the replacement model, namely the Adaptive Cell Under Test Estimator (ACUTE), is derived, 
allowing for the detection of small targets with adaptivity with respect to the background abundance 
estimated in the PUT. A modified version of the replacement model is developed in \cite{VINCENT2021108212}, 
where the GLRT is derived in the presence of a residual additive noise.

However, since in the hyperspectral sensors the spectra of different sub-pixel targets are mixed together with the background spectrum, a  {\em generalized replacement model}  is proposed in this paper,  where the sum of the 
total amount of both multiple sub-pixel targets and background spectra is equal to one, as explained ahead. 
In this way, the problem of detecting the presence of multiple sub-pixel targets is 
formulated as a binary hypothesis test where under the alternative hypothesis
the target is modelled in terms of a linear combination of endmembers whose coefficients also account
for the presence of the background. This model allows us
to detect and identify one or more targets from a wide spectral library of plausible targets, such as
different car types in a parking area, or a single target characterized by multiple spectral 
signatures, such as the pickup truck not considered in \cite{8964590}.
The detection problem at hand is solved by deriving decision rules where 
the unknown parameters, the background statistics, and the abundance vector are 
replaced by suit\nobreak able estimates based upon available secondary data collected around the PUT. 
Particularly, an iterative approach is proposed for the estimation of the unknown abundance 
vector and two different solutions (heuristic and constrained solutions) are considered at this end.
Finally, it is worth noticing that, as a byproduct, the devised detection architectures allow identifying the specific sub-pixel targets  in the PUT, from the spectral library 
of possible endmembers, by exploiting their corresponding estimated abundances.

The remainder of the paper is organized as follows. Section \ref{problem} is devoted 
to the replacement model and the formal statement of the detection problem. 
Two detection architectures are derived in Section \ref{detector}, which differ for the estimation 
of the target abundances. In Section \ref{results}, the behavior of the proposed 
architectures is investigated by means of both simulated as well as real data. 
Finally, concluding remarks end this article in Section \ref{conclusions}. 
Some derivations are confined to the appendices.

\section*{Notation}\label{Notation}
Vectors and matrices are denoted by boldface lower-case and upper-case letters, respectively.
Symbols $\det(\cdot)$ and $\Tr(\cdot)$ denote the determinant and the trace of a square matrix, 
respectively. Symbols $\bI$ and
$\mathbf{0}$ represent the identity matrix and the null vector or matrix of suitable 
dimensions, respectively. $\mathbf{1}$ is the vector of ones.
As to the numerical sets, $\R$ is the set of real numbers,
$\R^{N\times M}$ is the Euclidean space of $(N\times M)$-dimensional real matrices (or vectors if $M=1$).
We use $(\cdot)^T$ to denote the transpose while $\|\cdot\|$ is the Euclidean norm of a vector. 
The acronym PDF stands for Probability Density Function.
Finally, we write $\bx \sim \mathcal{N}_N  (\bmu, \bM)$ if $\bx$ is a $N$-dimensional Gaussian vector with
mean $\bmu\in\R^{N\times 1}$ and positive definite covariance matrix $\bM\in\R^{N\times N}$.
\section{Problem Statement}\label{problem}
This section defines a generalization of the so called replacement model \cite{8964590} 
that will be used to perform the detection in our case.
To this end, let us consider a hyperspectral sensor able to collect the 
reflected light (i.e., radiance) from the observed scene through a 
large number, say $N$, of spectral bands. The radiance  is generally  converted  
into  a reflectance spectrum to remove the effects of the non-uniform sun 
power-spectral density and the atmospheric contribution \cite{8738023,10.1117/12.816917}. 
The observed reflectance data samples from a given pixel can be grouped to form 
an $N$-dimensional vector, namely, 
\[
\by = \left[y_1,y_2, \ldots, y_N \right]^T\in \R^{N\times1}.
\]
In this work, a generalization of the replacement model \cite{10.1117/12.816917} 
is adopted, in which the presence of multiple sub-pixel targets (or otherwise stated endmembers) is supposed. 
The spectrum of each pixel can be expressed as a linear combination of $r$ endmembers plus the background component (that is, non-target)
\begin{equation}
    \label{rep-model}
    \by=\bT\balpha + \left(1-\balpha^T\mathbf{1}\right)\bb,
\end{equation}
where:
\begin{itemize}
\item $\bT = [\bt_{1}, \ldots, \bt_{r} ]\in\mathbb{R}^{N\times r}$ denotes the endmember matrix 
(the columns are their spectral signatures);
\item $\balpha=\left[\alpha_1,\ldots,\alpha_r\right]^T\in\mathbb{R}^{r\times 1}$ is the vector of the 
unknown fill factors subject to the constraints 
\begin{equation}
\alpha_i\geq0, \ \forall i \in \{1,\ldots,r\} \ \mbox{ and } \ \sum\limits_{i=1}^{r}\alpha_i<1;
\label{add}
\end{equation}
\item $\bb\in\mathbb{R}^{N\times1}$ is the background spectral signature.
\end{itemize}

This model is reasonable because it is likely for a pixel to comprise one or more target materials, due to the low spatial resolution of the majority of hyperspectral sensors.
The spectra of the different sub-pixel targets are mixed together (weighted by their 
respective abundances or fill factors) and with the spectrum of the background. 
As the abundances represent the proportion of the corresponding endmembers, 
the amount of both the sub-pixel targets and the background spectra is subject to the constraint 
that their sum is one.
A special case of \eqref{rep-model} is when $\alpha_i=0$, $\forall i \in \{1,\ldots,r\}$, 
which implies the absence of targets, i.e., the presence of background only. 
On the contrary, due to the strict inequality in \eqref{add}, the model does not admit the complete absence of the background's component. 

The detection problem aims at choosing between the null hypothesis 
$\cH_0$ ($\alpha_i=0, \forall i \in \{1,\ldots,r\}$) and the alternative 
hypothesis $\cH_1$ ($\alpha_i\neq0$, for at least one $i \in \{1,\ldots,r\}$). 
\begin{figure*}
\begin{equation}
    \label{f_H0}
    f_0(\by,\bZ;\bmu,\bM) = \left(\frac{1}{(2\pi)^{N/2}\det(\bM)^{1/2}}\right)^{K+1} \exp\left\lbrace  -\frac{1}{2}\Tr\left[ \bM^{-1}\left( (\by-\bmu)(\by-\bmu)^T +\sum_{k=1}^{K} (\bz_k - \bmu)(\bz_k - \bmu)^T\right) \right] \right\rbrace.
\end{equation}
\begin{align}\label{f_H1}
& f_1(\by,\bZ;\bmu,\bM,\balpha)=
 \left(\frac{1}{(2\pi)^{N/2}\left(1-\balpha^T\mathbf{1}\right)^{N/(K+1)}\det(\bM)^{1/2}}\right)^{K+1}
\nonumber\\
& \times \mathrm{exp}\left\lbrace 
-\frac{1}{2}\Tr\left[\bM^{-1}\left(
\frac{\left(\by\!-\!\bT\balpha\!-\!\left(1-\balpha^T\mathbf{1}\right)\!\bmu\right)
\left(\by\!-\!\bT\balpha\!-\!\left(1-\balpha^T\mathbf{1}\right)\!\bmu\right)^T}
{\left(1-\balpha^T\mathbf{1}\right)^2}
+\sum_{k=1}^{K} \left(\bz_k - \bmu\right)\left(\bz_k - \bmu\right)^T 
\right)\right]
\right\rbrace.
\end{align}
\end{figure*}
Using this system model, we can express our problem as the following binary hypothesis test
 \begin{equation}\label{Eq-HypothesisProblem}
 \left\{
    \begin{array}{l}
    \displaystyle

    \cH_0:
  \left\{
        \begin{array}{l}
        \displaystyle
        \by =\bb, \\
        \displaystyle
        \bz_k =\bb_k ,\quad k=1,\dots,K,
       \end{array}
   \right.

        \\
        \\

  \displaystyle
 \cH_1:
  \left\{
        \begin{array}{l}
        \displaystyle
        \by =\bT\balpha + \left(1-\balpha^T\mathbf{1}\right)\bb, \\
        \displaystyle
        \bz_k =\bb_k ,\quad k=1,\dots,K,
       \end{array}
   \right.

   \end{array}
 \right.
 \end{equation}
with $\bb$ and $\bb_k  \sim \mathcal{N}_N(\bmu,\bM)$. It is also supposed that $\bb$ and $\bb_k$ are statistically independent.
Notice that the detection problem presupposes that the background power varies between the two competing hypotheses.

Before concluding this section, we provide some definitions that will come in handy for 
the ensuing developments. More precisely, let $\bZ=\left[\bz_1,\,\dots,\bz_K\right]$ be the 
secondary data matrix, whose columns, $\bz_k$, $k=1,\dots,K> N$, 
are assumed to be statistically independent and identically distributed.
In this respect, the joint PDF of $\by$ and $\bZ$ under $\cH_0$ 
and $\cH_1$ can be expressed as in \eqref{f_H0} and \eqref{f_H1}, respectively.

\section{GLRT-based Detector Designs}
\label{detector}
In this section, we design decision rules for problem \eqref{Eq-HypothesisProblem} that are based upon 
the GLRT. Specifically, we modify this design procedure by exploiting suitable estimates for $\balpha$
that are different from the Maximum Likelihood Estimate (MLE). This choice is dictated by the difficult
mathematics arising from the application of the maximum likelihood approach to the estimation of $\balpha$
as required by the GLRT criterion.
Therefore, we start from the general equation of the GLRT, that is
\begin{align}\label{eq:LRT}
\frac{
\dmax_{\bmu,\bM,\balpha}f_1(\by,\bZ;\bmu,\bM,\balpha)}
{ \dmax_{\bmu,\bM} f_0(\by,\bZ;\bmu,\bM)}
\test \eta,
\end{align}
where $\eta$ is the detection threshold\footnote{Hereafter, we use symbol
$\eta$ to denote the generic threshold.} set according to
a given probability of false alarm (or probability of type I error), 
and proceed by separately solving the two optimization problems.

Under $\cH_0$, the problem at hand is well-known and, hence, for brevity, we show 
below the final results only. The MLEs of $\bmu$ and $\bM$ are given by
$\widehat{\bmu}_0 =\frac{1}{K+1} (\by + \tilde{\bz})$ with $\tilde{\bz}=\sum_{k=1}^{K}\bz_k$ and 
\begin{equation}
\widehat{\bM}_0 = \frac{\left[(\by-\widehat{\bmu}_0)(\by-\widehat{\bmu}_0)^T 
+\sum_{k=1}^{K} (\bz_k - \widehat{\bmu}_0)(\bz_k - \widehat{\bmu}_0)^T \right]}
{K+1},
\end{equation}
respectively, and
the final compressed log-likelihood under $\cH_0$ 
is\footnote{For simplicity, in what follows, we omit the dependence of the log-likelihood function on data $\by$ and $\bZ$.}
\begin{equation}
    \label{L0compr}
    L_0(\widehat{\bmu}_0,\widehat{\bM}_0) = - C_1 - C_2 \log{\det(\widehat{\bM}_0)} -N C_2,
\end{equation}
where $C_1=[(K+1)N/2]\log{(2\pi)}$ and $C_2={(K+1)}/{2}$.

Now, we focus on the $\cH_1$ hypothesis and write the corresponding log-likelihood (see \eqref{f_H1})
\begin{align}\label{f1_log}
&L_1(\bmu,\bM,\balpha)=
-C_1-N\log{A}-C_2\log{\left(\det{\bM}\right)} \nonumber
\\
&-\frac{\|\bM^{-1/2} \left(\bx-A\bmu\right) \|^2}
{2A^2} -\sum\limits_{k=1}^{K} 
\frac{\| \bM^{-1/2}\left(\bz_k - \bmu\right) \|^2}
{2},
\end{align}
where $A=\left(1-\balpha^T\mathbf{1}\right)$ and $\bx=\by-\bT\balpha$. 
We first maximize $L_1(\bmu,\bM,\balpha)$ with respect to $\bmu$, by setting to zero the corresponding derivative and obtain 
\begin{align}
&-\frac{1}{A}\bM^{-1}\bx +\bM^{-1}\bmu+\sum\limits_{k=1}^{K} \left(-\bM^{-1}\bz_k + \bM^{-1}\bmu\right)= 0
\\
& \Rightarrow  \left(K+1\right)\bM^{-1}\bmu=\frac{1}{A}\,\bM^{-1}\bx+\bM^{-1}\tilde{\bz}
\\
& \Rightarrow \widehat{\bmu}=\frac{1}{K+1}\left(\frac{1}{A}\,\bx+\tilde{\bz}\right).
\end{align}
Using the above results in \eqref{f1_log}, after some algebraic manipulations, 
the partially-compressed log-likelihood can be recast as
\begin{align}\label{eq:L_1maxMu}
&L_1(\boldsymbol{\widehat{\mu}},\bM,\balpha)=-C_1-N\log{A}-C_2\log{\left(\det{\bM}\right)} \nonumber
\\
&-\Tr\left\lbrace \frac{\bM^{-1}}{2} \left[\left(\frac{1}{A}\bx-\widehat{\bmu}\right)
\left(\frac{1}{A}\bx-\widehat{\bmu}\right)^{T}\!\!\!\!\!
+\sum\limits_{k=1}^{K} \left(\bz_k - \widehat{\bmu}\right)\left(\bz_k - \widehat{\bmu}\right)^{T}
\right]\right\rbrace\,.
\end{align}
The MLE of $\bM$ under $\cH_1$ can be computed by resorting to the following inequality \cite{lutkepohl1997handbook}
$\log \det(\bA) \leq \tr[\bA] - N$, where $\bA$ is any $N$-dimensional matrix with nonnegative eigenvalues, and, hence, 
we come up with
\begin{align}
\widehat{\bM}=\frac{
\left[\left(\frac{1}{A}\bx-\widehat{\bmu}\right)\!
\left(\frac{1}{A}\bx-\widehat{\bmu}\right)^{T}
+\sum\limits_{k=1}^{K} \left(\bz_k - \widehat{\bmu}\right)\left(\bz_k - \widehat{\bmu}\right)^{T}\right]
}
{K+1}\,.
\end{align}
Hence, we update  \eqref{eq:L_1maxMu} with   $\widehat{\bM}$ and find 
\begin{align}
\label{eq:beforeAppendixA}
&L_1(\boldsymbol{\widehat{\mu}},\widehat{\bM},\balpha)=-C_3-N\log{A}-C_2\log \det
\left[\left(\frac{1}{A}\bx-\widehat{\bmu}\right)
\right.
\nonumber
\\
&\left.\times \left(\frac{1}{A}\bx-\widehat{\bmu}\right)^{T}
+\sum\limits_{k=1}^{K} \left(\bz_k - \widehat{\bmu}\right)\left(\bz_k - \widehat{\bmu}\right)^{T}\right],
\end{align}
where $C_3=C_1+\frac{1}{2}\left(K+1\right)N-C_2N\log{\left(K+1\right)}$.
In Appendix \ref{appA}, we show that the argument of the determinant in \eqref{eq:beforeAppendixA}
can be suitably manipulated leading to the following expression for the partially-compressed
log-likelihood function
\begin{align}
&L_1(\boldsymbol{\widehat{\mu}},\widehat{\bM},\balpha)=-C_3-N\log{A} -C_2\log{\left(\det{\bS_1}\right)}-C_2
\nonumber
\\
&\times\log{\left(1+\frac{K}{K+1}\left(\frac{1}{A}\bx-\frac{1}{K}\tilde{\bz}\right)^T\!
\bS_{1}^{-1}\left(\frac{1}{A}\bx-\frac{1}{K}\tilde{\bz}\right)\right)}
\nonumber
\\
&=-C_3-N\log{\left(1-\balpha^T\mathbf{1}\right)} -C_2\log{\left(\det{\bS_1}\right)}
\nonumber
\\
&-C_2\log{\left(1+C_4
\left\|\bS_{1}^{-1/2}\left(\frac{\by-\bT\balpha}{1-\balpha^T\mathbf{1}}-\tilde{\tilde{\bz}}\right)\right\|^2\right)}\,,
\label{eq:afterAppendixA}
\end{align}
where $\bS_1=\bS-\frac{1}{K(K+1)}\tilde{\bz}\tilde{\bz}^T$ with $\bS=\bZ\bZ^T-\frac{1}{K+1}\tilde{\bz}\tilde{\bz}^T$,
$C_4=\frac{K}{K+1}$, and $\tilde{\tilde{\bz}}=\frac{1}{K}\tilde{\bz}$.
Since we are interested in the maximization of the partially-compressed 
log-likelihood with respect to $\balpha$, we focus on the terms that depend on $\balpha$ only and define the following
function
\begin{align}\label{eq:alphaterms}
&g(\balpha)=N\log{\left(1-\balpha^T\mathbf{1}\right)} \nonumber\\ 
&+C_2\log{\left[1+\left\|\left(\frac{\by_0-\bT_0\balpha}{1-\balpha^T\mathbf{1}}-\tilde{\tilde{\bz}}_0\right)^T\right\|^2
\right]}
\end{align}
where $\by_0=C_4^{1/2}\,\bS_{1}^{-1/2}\by$, $\bT_0=C_4^{1/2}\,\bS_{1}^{-1/2}\bT$ and $\tilde{\tilde{\bz}}_0=C_4^{1/2}\,\bS_{1}^{-1/2}\tilde{\tilde{\bz}}$.
The maximization of \eqref{eq:afterAppendixA} with respect to $\balpha$ is equivalent to the  problem 
\begin{equation}
\label{eq:problem}
  \left\{
        \begin{array}{l}
          \begin{aligned}
           \min_{\balpha} \quad &          g(\balpha) \\ 
           \textrm{subject to} \quad & \sum\limits_{i=1}^{r}\alpha_i<1,
           \\
           &\alpha_i\geq0, \quad \forall i \in \{1,\ldots,r\}.
          \end{aligned}
       \end{array}
   \right.
\end{equation}
In the next subsection, we describe two different procedures to solve problem \eqref{eq:problem}. Denoting
by $\widehat{\balpha}$ the generic solution returned by these procedures, we use it in \eqref{eq:afterAppendixA}
and the final expression of the detection architecture is
\begin{align}
    \label{finaldetector}
& L_1(\widehat{\bmu},\widehat{\bM},\widehat{\balpha}) - L_0(\widehat{\bmu}_0,\widehat{\bM}_0)
\test \eta.
\end{align}


\subsection{Solution to Equation \eqref{eq:problem}}
\label{subsec1}
The approach devised here relies on an iterative 
solution of \eqref{eq:problem}. In particular, we firstly highlight the dependence 
of the objective function from a single entry of $\balpha$, say $\alpha_j$, and
then, at each iteration, we minimize $g(\balpha)$ with respect to $\alpha_j$ as the index $j$ varies. 
To this end, let us notice that
\begin{equation}
\label{eq:firstAddendApp1}
1-\balpha^T\mathbf{1}
=1- \sum\limits_{i=1}^{r}\alpha_i
=1- \sum\limits_{i\neq j}\alpha_i-\alpha_j
=a_j-\alpha_j
\end{equation}
where $a_j=1- \sum\limits_{i\neq j}\alpha_i$ with $0<a_j<1$. Moreover, we have that
\begin{align}
\label{eq:secondAddendApp1}
\by_0-\bT_0\-\balpha
&=\by_0-\left[\bt_{0 1},\dots, \bt_{0 N}\right]\-\balpha
\nonumber
\\ 
&
=\by_0-\sum\limits_{i\neq j}\bt_{0 i}\,\alpha_i-\bt_{0 j}\,\alpha_j=\by_j-\bt_{0 j}\,\alpha_j
\end{align}
with $\by_j=\by_0-\sum_{i\neq j}\bt_{0 i}\,\alpha_i$.
The estimation procedure iterates according to the following rationale.
Denoting by $t$ the iteration index and given the estimates $\alpha_i^{(t+1)}$ (at the $(t+1)$th iteration),
$i=1,\ldots,j-1$, and $\alpha_i^{(t)}$ (at the $t$th iteration), $i=j+1,\ldots,r$,
we exploit $g(\balpha)$ to build up the following function of $\alpha_j$
\begin{align}\label{eq_derivate_alphaj}
&g(\alpha_j)=N\log{\left(\widehat{a}_j^{(t,t+1)}-\alpha_j\right)}+C_2
\nonumber
\\
&\times\log\left[1+\left\|\left(\frac{\widehat{\by}_{j}^{(t,t+1)}-\bt_{0 j}\alpha_{j}}
{\widehat{a}_j^{(t,t+1)}-\alpha_j}-\tilde{\tilde{\bz}}_0\right)\right\|^2\right],
\end{align}
where $\widehat{a}_j^{(t,t+1)}=1-\sum_{i=1}^{j-1}\alpha_i^{(t+1)}-\sum_{i=j+1}^{r}\alpha_i^{(t)}$
and $\by_j^{(t,t+1)}=\by_0-\sum_{i=1}^{j-1}\bt_{0 i}\,\alpha_i^{(t+1)}-\sum_{i=j+1}^{r}\bt_{0 i}\,\alpha_i^{(t)}$.
This function is then used to come up with the update of the estimate of $\alpha_j$ at the $(t+1)$h iteration.
Specifically, in the next subsections, we devise two different approaches: the first is heuristic whereas the second
 incorporates the constrained solutions  of \eqref{eq:problem} at the design stage.
An initial estimate of $\alpha_i, i=1,\dots,r$ is  necessary to initialize the 
algorithms as well as a reasonable stopping
criterion as, for instance, setting a maximum number of iterations, say $N_{iter}$.

\subsubsection{Heuristic solution}\label{Heur}
Let us recast \eqref{eq_derivate_alphaj} as
\begin{align}\label{eq_derivate_alphaj2}
&g(\alpha_j)=N\log{\left(\widehat{a}_j^{(t,t+1)}-\alpha_j\right)}+C_2\\\nonumber
&\times\left[\log{(D_0+D_1\alpha_j+D_2\alpha_j^2)} -2\log{\left(\widehat{a}_j^{(t,t+1)}-\alpha_j\right)} \right],
\end{align}
where $D_0=(\widehat{\by}_{j}^{(t,t+1)})^T\widehat{\by}_{j}^{(t,t+1)} - 2 \widehat{a}_j^{(t,t+1)} (\widehat{\by}_{j}^{(t,t+1)})^T 
\tilde{\tilde{\bz}}_0 +(\widehat{a}_j^{(t,t+1)})^2 \tilde{\tilde{\bz}}_0^T \tilde{\tilde{\bz}}_0 +(\widehat{a}_j^{(t,t+1)})^2$;
$D_1=2[ \widehat{a}_j^{(t,t+1)} (\bt_{0 j}^T \tilde{\tilde{\bz}}_0 
- \tilde{\tilde{\bz}}_0^T \tilde{\tilde{\bz}}_0)+(\widehat{\by}_{j}^{(t,t+1)})^T \tilde{\tilde{\bz}}_0 
-(\widehat{\by}_{j}^{(t,t+1)})^T  \bt_{0 j}] - 2 \widehat{a}_j^{(t,t+1)}$;
and $D_2 = 1+ \bt_{0 j}^T \bt_{0 j} -2 \bt_{0 j}^T \tilde{\tilde{\bz}}_0 + \tilde{\tilde{\bz}}_0^T \tilde{\tilde{\bz}}_0$.


Setting to zero the first derivative of \eqref{eq_derivate_alphaj2} with respect to $\alpha_j$ 
leads to the following quadratic equation
\begin{align}
    \label{alphajsol1}
    -ND_2 \alpha_j^2 + [2 C_2 D_2 \widehat{a}_j^{(t,t+1)}  + (C_2 - N) D_1] \alpha_j 
\nonumber
    \\
    + [C_2 D_1 \widehat{a}_j^{(t,t+1)} + 2  D_0 C_2 - N D_0] = 0.
\end{align}
Now, we can evaluate $\tilde{\alpha}_j^{(t+1)}$ by choosing the positive real-valued solution of \eqref{alphajsol1} 
returning the minimum value of \eqref{eq_derivate_alphaj2}. 
However, since the constraint \eqref{add} must be satisfied, we 
regularize $\tilde{\alpha}_i^{(t+1)}$, $i=1,\dots,r$, as follows
\begin{equation}
\label{normalization}
\widehat{\alpha}_i^{(t+1)}=\tilde{\alpha}_i^{(t+1)}
\frac{\left(1-\alpha_b \right)}{\sum\limits_{i=1}^{r}\tilde{\alpha}_i^{(t+1)}},
\end{equation}
where $0\leq\alpha_b<1$ represents 
the unknown background abundance; in practice, it can be set using a 
linear grid of values (sized according to the available a priori information) 
and selecting the value that minimizes the objective function.
This heuristic algorithm is summarized in Algorithm \eqref{algoHeur}.

\begin{algorithm}[htb!]
\caption{Estimation Procedure for $\alpha_j$ (heuristic solution)}
\label{algoHeur}
\begin{algorithmic}[1]
\REQUIRE $\bT_0$, $\by_0$, $\tilde{\tilde{\bz}}_0$ $\alpha^{(0)}_i,i=1,\dots,r$, $N_{iter}$
\ENSURE $\boldsymbol{\widehat{\alpha}}$
\STATE Set $t=1$
\STATE Set $j=1$
\STATE Compute $\ds\widehat{a}_j^{(t-1,t)}= 1-\sum_{i=1}^{j-1}\alpha_i^{(t)}-\sum_{i=j+1}^{r}\alpha_i^{(t-1)}$
\STATE Select the $j$th column of $\bT_0$, i.e., $\bt_{0 j}$
\STATE Compute $\ds\by_j^{(t-1,t)}=\by_0-\sum_{i=1}^{j-1}\bt_{0 i}\,\alpha_i^{(t)}-\sum_{i=j+1}^{r}\bt_{0 i}\,\alpha_i^{(t-1)}$
\STATE Compute $\tilde{\alpha}_j^{(t)}$ by solving \eqref{alphajsol1} and selecting the positive real-valued solution 
that minimizes \eqref{eq_derivate_alphaj2}
\STATE If  $j< r$, set $j=j+1$ and go to step $3$ else go to step $8$
\STATE Normalize $\tilde{\balpha}^{(t)}=[\tilde{\alpha}_1^{(t)},\dots,\tilde{\alpha}_r^{(t)}]^T$  as in \eqref{normalization}
to obtain $\widehat{\balpha}^{(t)}=[\widehat{\alpha}_1^{(t)},\dots,\widehat{\alpha}_r^{(t)}]^T$
\STATE If $t< N_{iter}$, set $t=t+1$ and go to step $2$ else go to step $10$
\STATE Return $\widehat{\balpha}=[\widehat{\alpha}_1^{(t)},\dots,\widehat{\alpha}_r^{(t)}]^T$
\end{algorithmic}
\end{algorithm}

\subsubsection{Constrained solutions}
\label{Constr}
Let us introduce an auxiliary variable, say $\beta_j$, such that 
\begin{align}
&\beta_j+\sum_{i=1}^{j-1}\alpha_i^{(t+1)}+\sum_{i=j+1}^{r}\alpha_i^{(t)}+\alpha_j=1
\\
&\Rightarrow 
\beta_j+\alpha_j=\widehat{a}_j^{(t,t+1)}.
\end{align}
Then, we exploit $\beta_j$ to modify \eqref{eq_derivate_alphaj} by incorporating 
the model constraint on the abundances, namely
\begin{align}
    \label{Lbeta}
    & g(\alpha_j,\beta_j)=N\log{\beta_j} +C_2 
    \\ 
&\times\log\left[1+\left\|\left(\frac{\widehat{\by}_{j}^{(t,t+1)}-\bt_{0 j}\alpha_{j}}{\beta_j}
-\tilde{\tilde{\bz}}_0\right)\right\|^2\right], \nonumber
\end{align}
and consider the following minimization problem
\begin{equation}
\label{lagrange}
\left\{  
\begin{array}{l}
\begin{aligned}
&\min_{\alpha_j,\beta_j} g(\alpha_j,\beta_j) \\
&\text{subject to } \alpha_j + \beta_j =  \widehat{a}_j^{(t,t+1)}\\
\end{aligned}
\end{array}
\right. .
\end{equation}
Now, we apply the method of Lagrange multipliers and define the Lagrangian
\begin{equation}
    \label{L_lagrange}
    \cL(\alpha_j,\beta_j) = g(\alpha_j,\beta_j) - \lambda \left( \alpha_j + \beta_j - \widehat{a}_j^{(t,t+1)} \right),
\end{equation}
where $\lambda$ is a Lagrange multiplier.
Setting to zero the gradient of the Langrangian and considering the constraint equation,
we form the following system of equations
\begin{equation}
    \label{LagrangeProblem}
    \left\{  
\begin{array}{l}
\begin{aligned}
&\lambda A_1 \alpha_j^2 + (\lambda A_2 - 2C_2 A_1) \alpha_j + \lambda A_3 - C_2 A_2 = 0 
\\ 
& -\lambda B_1 \beta_j^3 + (N B_1 - \lambda B_2) \beta_j^2 + (N B_2 - C_2 B_2 - \lambda B_3) \beta_j
\\
&+ N B_3 - 2 C_2 B_3 = 0 \\
&\alpha_j + \beta_j =  \widehat{a}_j^{(t,t+1)}
\end{aligned}
\end{array} \right.
\end{equation}
where $A_1=\bt_{0 j}^T \bt_{0 j}$; $A_2=2 [ \beta_j \bt_{0 j}^T \tilde{\tilde{\bz}}_0 - (\widehat{\by}_{j}^{(t,t+1)})^T  \bt_{0 j}]$;
$A_3= \beta_j^2 (1+\tilde{\tilde{\bz}}_0^T \tilde{\tilde{\bz}}_0) -2 \beta_j (\widehat{\by}_{j}^{(t,t+1)})^T  \tilde{\tilde{\bz}}_0 
+ (\widehat{\by}_{j}^{(t,t+1)})^T \widehat{\by}_{j}^{(t,t+1)}$; $B_1=(1+\tilde{\tilde{\bz}}_0^T \tilde{\tilde{\bz}}_0)$;
$B_2 = 2 [\alpha_j \bt_{0 j}^T \tilde{\tilde{\bz}}_0 - (\widehat{\by}_{j}^{(t,t+1)})^T \tilde{\tilde{\bz}}_0 ] $;
$B_3 = (\widehat{\by}_{j}^{(t,t+1)})^T \widehat{\by}_{j}^{(t,t+1)} - 
2 \alpha_j (\widehat{\by}_{j}^{(t,t+1)})^T \bt_{0 j} + \alpha_j^2 \bt_{0 j}^T \bt_{0 j}$.

Finally, the estimate of $\widehat{\alpha}_j$, say $\widehat{\alpha}_j^{(t+1)}$, 
is obtained by selecting the real-valued positive solution that is strictly lower than 1 and minimizes 
the objective function as summarized in Algorithm \eqref{algoConstr}.


\begin{algorithm}[htb!]
\caption{Estimation Procedure for $\alpha_j$ (constrained solution)}
\label{algoConstr}
\begin{algorithmic}[1]
\REQUIRE $\bT_0$, $\by_0$, $\tilde{\tilde{\bz}}_0$ $\alpha^{(0)}_i,i=1,\dots,r$, $N_{iter}$
\ENSURE $\boldsymbol{\widehat{\alpha}}$
\STATE Set $t=1$
\STATE Set $j=1$
\STATE Compute $\ds\widehat{a}_j^{(t-1,t)}= 1-\sum_{i=1}^{j-1}\alpha_i^{(t)}-\sum_{i=j+1}^{r}\alpha_i^{(t-1)}$
\STATE Select the $j$th column of $\bT_0$, i.e., $\bt_{0 j}$
\STATE Compute $\ds\by_j^{(t-1,t)}=\by_0-\sum_{i=1}^{j-1}\bt_{0 i}\,\alpha_i^{(t)}-\sum_{i=j+1}^{r}\bt_{0 i}\,\alpha_i^{(t-1)}$
\STATE Compute $\widehat{\alpha}_j^{(t)}$ by solving \eqref{LagrangeProblem} and selecting 
the real-valued positive solution strictly that is lower than $1$ and minimizes \eqref{Lbeta}
\STATE If  $j<r$, set $j=j+1$ and go to step $3$ else go to step $8$
\STATE If $t<N_{iter}$ set $t=t+1$ and go to step $2$ else go to step $9$
\STATE Return $\widehat{\balpha}=[\widehat{\alpha}_1^{(t)},\dots,\widehat{\alpha}_r^{(t)}]^T$
\end{algorithmic}
\end{algorithm}

\section{Performance Analysis}\label{results}


\begin{figure*}[htb] \centering
	\includegraphics[width=2\columnwidth]{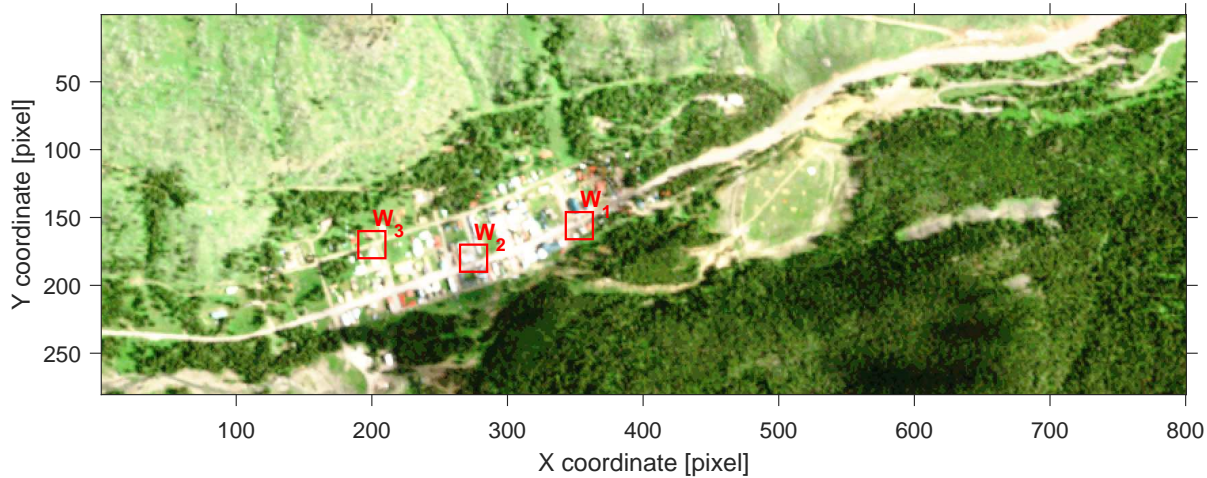}
	\caption{Cooke City scene of the RIT dataset in RGB representation 
	combining bands 15, 8, and 3 respectively. $W_1$, $W_2$ and $W_3$ represent the three test windows used to perform the multiple sub-pixel target analysis described in the following sections.}
	\label{fig:RIT_image}
\end{figure*}

In  this section, we assess the detection performance of the proposed detectors through numerical examples
based on simulated as well as real data. 
To this end, we resort to an hyperspectral dataset, namely the Rochester Institute 
of Technology (RIT) experiment\footnote{Data can be downloaded from  \url{http://dirsapps.cis.rit.edu}} \cite{4779144}.
The RIT open data experiment has been specially designed for target detection 
and has been widely used in the open literature \cite{8964590,doi:10.1080/22797254.2020.1850179}. 
Indeed, a corrected and geo-registered reflectance map is available so that the detection 
performance will be independent from any particular experimental setup.

Data were collected in July 2006 with a coverage area of 
approximately 2.0 km$^2$ and around the small town of Cooke City, Montana, USA. 
To this end, the airborne HyMap sensor operated by HyVista was used \cite{Cocks19981TH}. 
The images were acquired flying at $1.4$ km above the ground and were successively 
geo-registered using ground control points. Both calibrated spectral radiance as well as 
spectral reflectance after atmospheric compensation are available in the dataset.

The Cooke City scene is shown in Figure~\ref{fig:RIT_image}, which is composed by $280 \times 800$ pixels. Each pixel is observed at 126 spectral bands covering the electromagnetic spectrum from 0.45 $\mu$m to 2.48 $\mu$m with a ground resolution of about $3.0 \times 3.0$ m. It is important to note that the spatial resolution of the map is of the same order of magnitude as the target sizes, so that they will usually behave as sub-pixel targets \cite{10.1117/12.849910}, \cite{Khazai}. 

\begin{figure}[htb] \centering
	\subfigure[Vehicles spectral signatures]{\includegraphics[width=0.8\columnwidth]{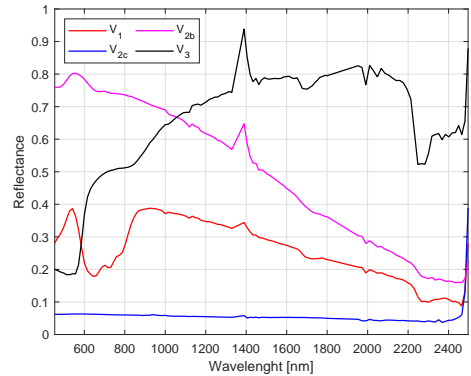}}
	\subfigure[Fabric panels spectral signatures]{\includegraphics[width=0.8\columnwidth]{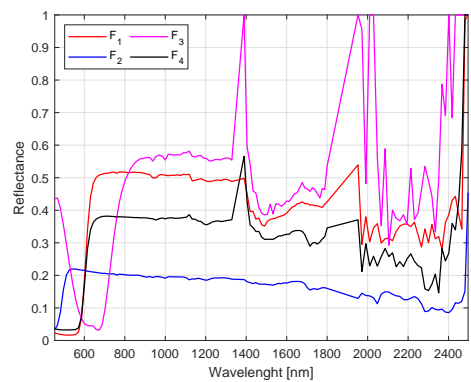}}
	\caption{Targets spectral signatures of RIT dataset.}
	\label{fig:Spectral_signatures}
\end{figure}

In this dataset, civilian vehicles and small fabric panels were used as targets. Specifically, three kinds of cars (indicated as V$_1$, V$_2$ and V$_3$) and four different fabric panels (F$_1$, F$_2$, F$_3$ and F$_4$) are present in the scene. It is important to highlight that V$_2$ is a pick-up characterized by two different spectral signatures, namely, one corresponding to the cabin (V$_{2c}$) and the other to the back (V$_{2b}$), so it can be considered as an example of multi-target. 
For each target, a reference spectrum signature obtained from a laboratory spectrophotometer is provided together with the RIT dataset, as shown in Figure \ref{fig:Spectral_signatures}. 

Finally, the RIT dataset provides a standard self-test where 
 the targets' map positions  are known, and also  a blind test with unknown target positions to prevent ad hoc algorithms. 
Moreover, the water absorption and low signal-to-noise bands were 
identified and removed from the Cooke City dataset for further processing. 
Precisely, bands no. 1, 2, 3, 63, 64, 65, 66, 95, 96, and 97 were discarded 
as in \cite{985871}. After removing these bands, 116 spectral bands were retained. 

\subsection{Results on simulated data}

In this subsection, a reflectance pixel containing the target vehicle V$_2$ 
is simulated according to the replacement model defined in (\ref{rep-model}). 
Particularly, the considered endmembers' matrix is composed by three spectral signatures
$$\bT = [\bt_{2c}, \bt_{2b},\bt_{3} ]\in\mathbb{R}^{116\times 3},$$ 
where $\bt_{2c}$, $\bt_{2b}$, and $\bt_{3}$ denote the 
spectral signatures of V$_{2c}$, V$_{2b}$, and V$_3$, respectively, 
that are given together with the RIT dataset. 

We consider different configurations for the fill factor 
vector of the abundances, as specified in Table \ref{Tab:tabella_fill}. 
It is important to note that for the  V$_{2c}$ endmember we assign a bigger value of abundance as its reflectance signature is lower if compared with V$_{2b}$, see Figure \ref{fig:Spectral_signatures}(a). 
We add a background noise modelled in terms of a zero-mean Gaussian random vector with variance 0.5 and independent entries. The number of secondary data $K$ is set to 625.

Remember that the heuristic approach, as specified in \eqref{normalization}, requires a selection of a linear grid of values for the background abundance to minimize the objective function. Thus, we set a linear grid of values from $0.1$ to $0.9$, with a step of $0.01$. This condition is applied for performances evaluated on both simulated and real data.

\begin{table}[htp]
\begin{center}
\caption{Abundances for different simulated pixel test cases. 
$\alpha_{2c}$, $\alpha_{2b}$, and $\alpha_{3}$ are the abundances 
referred to the endmembers V$_{2c}$, V$_{2b}$, and V$_{3}$, respectively.} 
\begin{tabular}{l | r | r}
	$(\alpha_{2c}, \alpha_{2b}, \alpha_{3})$  & abundances'  & background's   \\
		 & sum & abundance  \\
	\hline
	$(0.00, 0.00, 0)$ 		& $0.00$    & $1.00$	\\
	$(0.31, 0.01, 0)$ 		& $0.32$ 	   & $0.68$ \\
	$(0.32, 0.02, 0)$ 		& $0.34$ 	   & $0.66$\\
	$(0.33, 0.03, 0)$ 		& $0.36$ 	  	& $0.64$\\
	$(0.34, 0.04, 0)$ 		& $0.38$ 	  	& $0.62$\\
	$(0.35, 0.05, 0)$ 		& $0.40$ 	  	& $0.60$\\
	$(0.40, 0.10, 0)$ 		& $0.50$ 	  	& $0.50$\\
	$(0.45, 0.15, 0)$ 		& $0.60$ 	   & $0.40$\\
	$(0.50, 0.20, 0)$ 		& $0.70$ 	  	& $0.30$\\
	$(0.55, 0.25, 0)$ 		& $0.80$ 	  	& $0.20$\\
	$(0.60, 0.30, 0)$ 		& $0.90$ 	  	& $0.10$\\
\end{tabular}
\label{Tab:tabella_fill}
\end{center}
\end{table}%

\begin{figure}[!t] \centering
	\subfigure[Sum of abundances equal to 0.7]{\includegraphics[width=0.8\columnwidth]{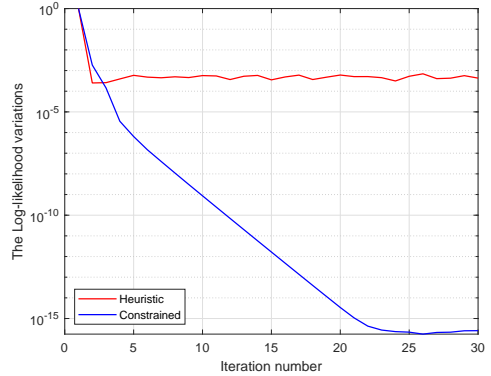}}
	\subfigure[Sum of abundances equal to 0.9]{\includegraphics[width=0.8\columnwidth]{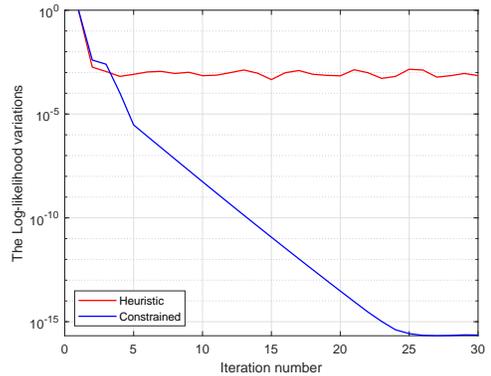}}
	\caption{Log-likelihood mean variation versus the iteration number over 100 MC independent trials.}
	\label{fig:convergence}
\end{figure}

As a preliminary step, we analyze the behavior of the proposed procedures in terms of the number 
of iterations required for convergence. To this end, we define the Log-likelihood variation $\Delta L_1^{(h)}$ as a function of the iteration index, say $h$, as
\begin{equation}
\label{eq:convergence}
\Delta L_1^{(h)}=\left|\frac{L_1^{(h)}(\widehat{\bmu},\widehat{\bM},\widehat{\balpha}) 
- L_1^{(h-1)}(\widehat{\bmu},\widehat{\bM},\widehat{\balpha})}{L_1^{(h)}(\widehat{\bmu},\widehat{\bM},\widehat{\balpha})}\right|. 
\end{equation}
In this analysis, two different cases are considered with abundances' sums 
of 0.7 and 0.9 (as in Table \ref{Tab:tabella_fill}), respectively.
In Figure \ref{fig:convergence}, we plot the average values of (\ref{eq:convergence}) 
evaluated over $100$ Monte Carlo (MC) independent trials for both the 
heuristic and the constrained approaches. From Figure \ref{fig:convergence}, 
it is clear that for the heuristic algorithm the Log-likelihood variation settles at an approximately constant value. In fact, after 5 iterations, $\Delta L_1^{(h)}$ varies in the interval from $10^{-4}$ to $10^{-3}$.
Contrarily, the constrained method requires more than 20 iterations 
to reaches a constant value below $10^{-15}$  for $\Delta L_1^{(h)}$. 
In the following, we will set the maximum number of iterations as $N_{iter}=15$ 
which is sufficient for obtaining $\Delta L_1^{(h)} < 10^{-2}$ for both heuristic and constrained approaches.

\begin{figure}[!t] \centering
	\subfigure[True abundances as shown in Table \ref{Tab:tabella_fill}]{\includegraphics[width=0.8\columnwidth]{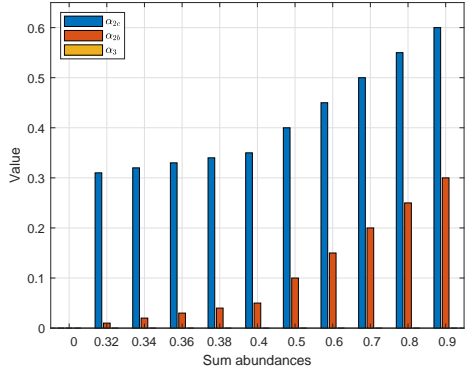}}
	\subfigure[Estimated abundances Heuristic method]{\includegraphics[width=0.8\columnwidth]{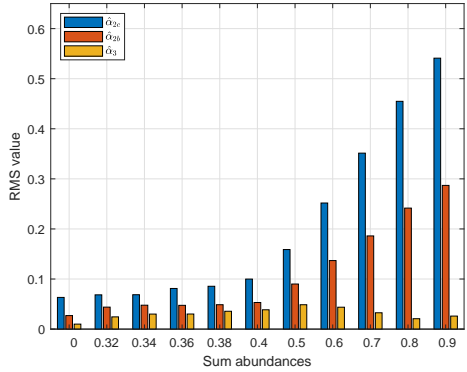}}
	\subfigure[Estimated abundances Constrained method]{\includegraphics[width=0.8\columnwidth]{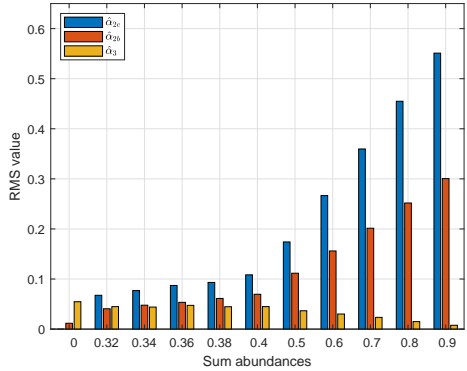}}
	\caption{RMS value of estimated abundances for the 
	heuristic and constrained approaches over $1000$ MC trials and for different background interval.}
	\label{fig: Stime_H_C}
\end{figure}

In Figure \ref{fig: Stime_H_C}(a) we plot the true abundance's configurations specified in Table \ref{Tab:tabella_fill} while in Figure \ref{fig: Stime_H_C}(b) and Figure \ref{fig: Stime_H_C}(c) we plot the Root Mean Square (RMS) value of the estimated abundances, averaged over 1000 MC trials, for both the heuristic and the constrained approaches, respectively. It is immediately evident that the estimate trends of the heuristic and the constrained methods are very similar. Both methods, in fact, for low concentrations of targets' abundances, i.e., high background concentration, provide estimate values that differ considerably from the true ones. This behavior 
can be explained if we look at the Table \ref{Tab:tabella_fill}, where when abundance sum is less 
than $0.50$, the single endmembers' abundances, i.e., $\alpha_{2c}$, $\alpha_{2b}$, and $\alpha_{3}$, 
are less than the background concentration and represent a very challenge situation. On the contrary, we notice that for abundances' sum greater than $0.50$, we come up with reasonable estimates of each target's abundance. Specifically, when abundance's sum is greater than or equal to $0.80$, the estimated values are very close to the true values. In these configurations, the background's concentration is less than each abundances' value.

\begin{figure}[!t] \centering
	\includegraphics[width=0.8\columnwidth]{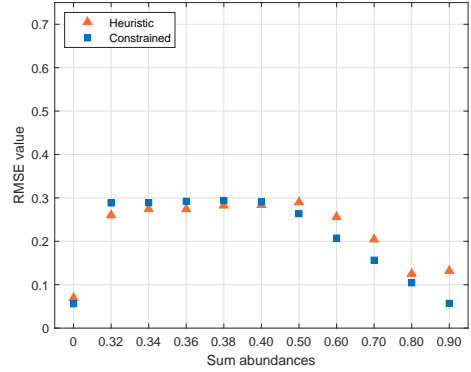}
	\caption{RMSE value for heuristic and constrained approach computer for 1000 MC trials.}
	\label{fig:RMSE_value}
\end{figure}

A more accurate analysis of the abundances' estimates obtained for both 
constrained and heuristic algorithms is performed in terms of RMS Error (RMSE), which is shown in
Figure \ref{fig:RMSE_value}. As reasonable to expect, the RMSE trend is the same for both algorithms. Specifically, from $0.32$ to $0.50$ of the sum abundances, the RMSE is almost constant and it presents the higher values. In this interval we note that the RMSE of heuristic approach is slightly less than that of the constrained one. For abundance' sum  of $0.5$, the RMSE begins to decrease linearly, confirming that estimated values are closer to the true ones. In this case the constrained approach provides better estimation performance than the heuristic method. 

\begin{figure}[htb] \centering
	\includegraphics[width=0.8\columnwidth]{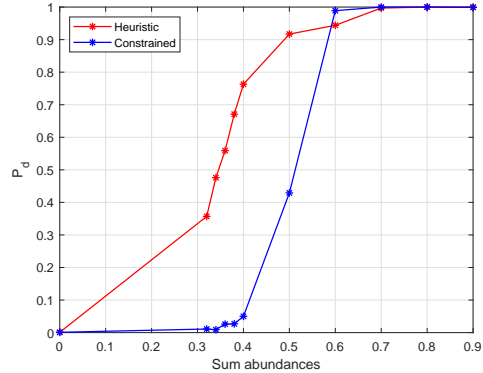}
	\caption{Detection probability computed for $P_{fa}=10^{-3}$ and for 1000 MC independent trials.}
	\label{fig:Probability_detection}
\end{figure}

Figure \ref{fig:Probability_detection} shows the detection probability $P_{d}$ evaluated 
using a false alarm probability $P_{fa}=10^{-3}$ and $1000$ MC trials. From this figure, 
it turns out that the $P_d$ values of the heuristic approach are higher than those of the constrained approach in the interval of abundances less than or equal to $0.5$. This trend, in accordance with what has already been said for the RMSE, can be due to better estimated values. 
For abundances' sum greater than $0.5$, the detection probabilities are greater than $0.9$ for both approaches. Particularly, we obtain the maximum value for the detection probability at $0.6$ and $0.7$ of abundance's sum for the constrained and the heuristic approaches, respectively.

\subsection{Results on real data}
In this section, the performance of the proposed architectures is assessed through the real RIT dataset.
This analysis allows us to quantify the robustness of the proposed detectors
in the presence of model mismatches due to the fact that real data do not exactly match
the design assumptions.

At first, we assess the performance in the case of a single sub-pixel target. 
To this aim, we consider a single signature and we use the ACUTE detector, recently proposed in \cite{8964590}, as competitor.
Next, the detection performance for multiple sub-pixel targets is analyzed.

It is important to highlight that no specific pre-processing has been applied to the real RIT dataset. 
Finally, for numerical reasons, we scale the reflectance spectral signature 
(shown in Figure \ref{fig:Spectral_signatures}) by a factor of 100.

~\\
\noindent
\textit{1) Single sub-pixel target detection}

\begin{figure}[!t] \centering
	\includegraphics[width=0.8\columnwidth]{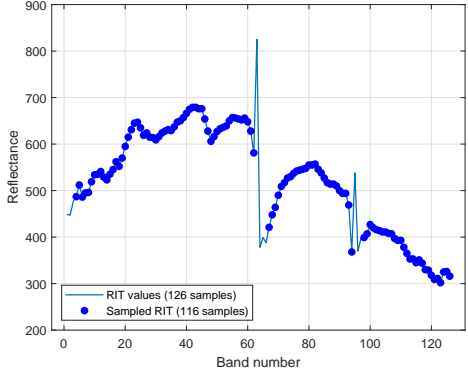}
	\caption{Spectral reflectance of target V$_3$ located at pixel $P_3 \equiv (282,186)$. Blue dots indicate the considered spectral samples to avoid water absorption and low SNR bands.}
	\label{fig:Reflectance_pixel_V3}
\end{figure}

The objective of this subsection is to compare the performance of the heuristic 
and constrained detectors with the ACUTE detector \cite{8964590}. To this end, 
we use the entire RIT dataset and as target of interest we consider V$_3$ only. 
The choice of the target V$_3$ is dictated by the fact that it is the most 
challenging in terms of false alarms, as shown in \cite{8964590}. The V$_3$ target, as indicated by 
the information related
to the dataset, has pixel coordinates: $P_3 \equiv (282,186)$. 
Figure \ref{fig:Reflectance_pixel_V3} shows the spectral reflectance for this target pixel.

In order to make a comparable performance analysis with the ACUTE detector, the spectral matrix is composed by only the spectral signature of target V$_{3}$: $\bT \equiv \bt_{3} \in\mathbb{R}^{116\times 1}$.

\begin{table}[!b]
	\begin{center}
		\captionof{table}{Abundances estimation of target V$_3$ at pixel $P_3 \equiv (282,186)$, for different background window size.} 
		\begin{tabular}{c | c c c }
			$K$				\quad	& \quad\quad \quad ACUTE 	  \quad		& Heuristic	     \quad   & Constrained \\
			\hline
			$ 15 \times 15 $ \quad	& \quad\quad \quad $0.003$ 		\quad		& $0.100$    \quad   & $0.051$    \\
			$ 25 \times 25 $  \quad	& \quad\quad  \quad $0.039$ 		\quad		& $0.100$    \quad   & $0.051$    \\
			$ 55 \times 55 $  \quad	&  \quad \quad \quad $0.142$ 			\quad	& $0.140$    \quad   & $0.136$    \\
		\end{tabular}
		\label{Tab:tabella_stime_V3}
	\end{center}
\end{table}

In Table \ref{Tab:tabella_stime_V3}, we report the abundance's estimate relative to the target V$_{3}$ 
for all the considered algorithms. These results are obtained by applying three background window sizes, 
namely, $15 \times 15$, $25 \times 25$, and $55 \times 55$ pixels around the PUT; moreover, a $3 \times 3$ 
pixels guard window is considered. We note that a small background window size, such as $15 \times 15$ or $25 \times 25$, 
results in low abundance estimates of the target, whereas the $55 \times 55$ window returns 
an higher abundance's estimation as well as comparable abundances between the three detectors.  
For this reason, in what follows, we select a background window of $55 \times 55$ pixels.



The detection performance is assessed in terms of false alarm rate.\footnote{Specifically, 
we repeat the same analysis conducted in \cite{8964590}, where the false alarm rate is evaluated as 
the number of no target pixels having their detector's statistic strictly higher than the one calculated using $P_3$.} 
Table \ref{Tab:tabella_performance} shows the false alarm rate achieved for the three detectors.  
From this analysis, we note that the false alarm rate of the heuristic and the constrained detectors 
are almost the same and both lower than that of the ACUTE detector.

\begin{table}[!t]
	\begin{center}
		\captionof{table}{Performance comparison between ACUTE and proposed detectors in term of false alarm rate for target V$_{3}$ of RIT dataset.} 
		\begin{tabular}{c | c c c }
			$K$				\quad	& \quad\quad \quad   ACUTE 	  \quad		& Heuristic	  \quad   & Constrained \\
			\hline
			$ 55 \times 55 $  \quad	&  \quad \quad \quad $6.728\%$ \quad   & $4.453\%$    \quad   & $4.483\%$    \\
		\end{tabular}
		\label{Tab:tabella_performance}
	\end{center}
\end{table}

~\\
\noindent
\textit{2) Multiple sub-pixel targets detection}

In this subsection, we consider the target V$_2$, a multiple target case as it is represented 
by the two spectral signatures: the cabin target V$_{2c}$ (with signature $\bt_{2c}$) and the 
back target V$_{2b}$ (with signature $\bt_{2b}$). 
Since for this scenario the ACUTE detector cannot be used, we will focus only on the results 
obtained through the heuristic and the constrained detectors and we use the ground truth from the dataset. 
As indicated by data description, the V$_2$ target is located at pixel coordinates $P_2\equiv(353,156)$. 

At first, we focus on the abundances' estimation for this pixel $P_2$. In this analysis, we consider 
two different configurations for the spectral matrix. Specifically, we take into account 
the spectral matrix already defined in the simulated scenario, i.e., 
$\bT= [\bt_{2c}, \bt_{2b},\bt_{3} ]\in\mathbb{R}^{116\times 3}$, and 
the spectral matrix made by the two spectral signatures of the V$_2$ target only, i.e., 
$\bar{\bT} = [\bt_{2c}, \bt_{2b}]\in\mathbb{R}^{116\times 2}$. 
Around the PUT, the background windows of size $ 55 \times 55$ pixels and the $3 \times 3$ pixels 
guard window are applied. Using both spectral matrices $\bar{\bT}$ and $\bT$, and inspecting the 
target abundance estimates for $\bt_{2c}$ and $\bt_{2b}$, 
i.e., $\widehat{\alpha}_{2c}$ and $\widehat{\alpha}_{2b}$, respectively, 
we obtain low values for both algorithms.
In particular, with focus on the V$_{2c}$ target,
the heuristic approach 
returns $\widehat{\alpha}_{2c} \approx 0.094$, while the value obtained by means of the constrained approach 
is $\widehat{\alpha}_{2c} \approx 0.024$. 
As for target V$_{2b}$, the estimated abundance is $\widehat{\alpha}_{2b} \approx 0.006$ for the 
heuristic approach and zero for the constrained one.
Even though the true abundance's value are not given in the dataset, 
the estimated abundances related to $P_2$ are low in spite of the claimed
presence of V$_{2}$ in that pixel. 
This situation is probably due to possible mismatches between the real 
target signature and the presumed one. 


In order to evaluate the detection performance in a multiple sub-pixel scenario, 
we consider the three test windows shown in Figure \ref{fig:RIT_image}, 
denoted by $W_1$, $W_2$, and $W_3$, and of size $21 \times 21$. Such windows are 
representative of different scenarios. Specifically, window $W_1$ is exactly centered where is 
located V$_2$ target, i.e, $P_{W_1} \equiv  P_{2} \equiv (353, 156)$, and is characterized 
by a mixed presence of vegetation and anthropic elements, such as roads, houses, and buildings. 
The second window, namely $W_2$, is centered on pixel $P_{W_2} \equiv (275, 180)$ and it mainly 
encloses an urban area. Finally, the $W_3$ window, centered on pixel $P_{W_3} \equiv (200, 170)$, 
contains low vegetation. Given the most uniform coverage of ${W_3}$, we 
assume that the pixels of this window represent background only. 
Therefore, we set the detection threshold over ${W_3}$ with
$P_{fa}=10^{-2}$. Specifically, the threshold value is estimated for 
each spectral matrix configuration, i.e., $\bar{\bT}$ and $\bT$, and
both approaches. 
Table \ref{Tab:tabella_pfa} summarizes the false alarm rates computed over the other two windows, 
namely, ${W_1}$ and ${W_2}$. 
It is immediately evident that the false alarm rates for the heuristic and constrained approaches are of the same order for each test window. Specifically, regardless of the spectral matrix applied, the false alarm rate is 
about $1\%$ for ${W_1}$ window and is about $6\%$ for the ${W_2}$ window. 
Notice that for the selected thresholds, target V$_{2}$, which is present in ${W_1}$, would not be detected. 
On the contrary, target V$_{3}$, located at pixel $P_3 \equiv (282,186)$, is within the ${W_2}$ window and a 
detection is obtained in its 3 $\times$ 3 pixels guard window, specifically at pixel with coordinates $(282, 185)$.
Finally, it is worth noticing that the high number of false alarm in $W_2$ might be due 
to the presence of several anthropic elements.

\begin{table}[!b]
	\begin{center}
		\captionof{table}{False alarm rate for both Heuristic and Constrained approaches over $W_1$ and $W_2$ windows applying two configurations of the spactral matrix.  } 
		\begin{tabular}{c | c c }
			Window				\quad	& \quad 		$\bar{\bT}$ 	  \quad		& $\bT$ \\
			\hline
			\multirow{2}{*}{$W_1$} \quad	& \quad			Heuristic: $0.907\%$  	  \quad   & Heuristic: $1.133\%$   \\
											& \quad 		Constrained: $1.133\%$\   \quad   & Constrained: $1.133\%$    \\
			\hline
			\multirow{2}{*}{$W_2$} \quad	& \quad			Heuristic: $5.895\%$  	  \quad   & Heuristic: $6.122\%$   \\
											& \quad			Constrained: $6.122\%$\     \quad   & Constrained: $6.122\%$    \\
		\end{tabular}
		\label{Tab:tabella_pfa}
	\end{center}
\end{table}

To further investigate the behavior of the proposed detectors, we fictitiously introduce the V$_{2}$ target within a real pixel of the RIT dataset. Specifically, we identify a background pixel that corresponds to $P'_2 \equiv (240,155)$, and according to the replacement model, we insert multi-target V$_{2}$ into the real pixel. Specifically, we denote by $\by_{F}(\balpha_n)$ the spectral reflectance values of the filled pixel, with $n$ indicating a generic filling configuration corresponding to background 
values ranging from 0.6 to 0.1, and we define it as
\begin{equation}
\label{filled}
\by_{F}(\balpha_n)=[\bt_{2c}, \bt_{2b}]\balpha_{n}+(1-\balpha_n^T\mathbf{1})\by_{RIT}, 
\end{equation}
where $\by_{RIT}$ is the pixel reflectance of the RIT dataset, and $\balpha_n=[\alpha_{2c},\alpha_{2b}]^T$ with $\alpha_{2c}$ and $\alpha_{2b}$ the abundances' values for V$_{2c}$ and V$_{2b}$, respectively. 

\begin{figure}[!t] \centering
	\includegraphics[width=0.8\columnwidth]{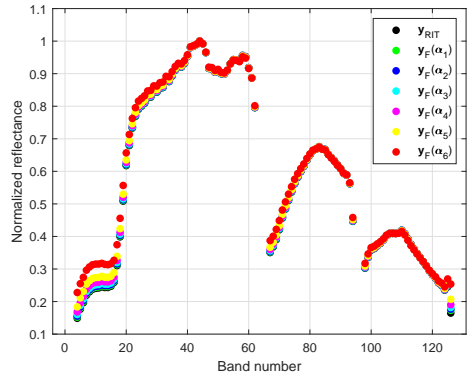}
	\caption{Spectral reflectance normalized to the maximum value of the considered 116 spectral samples of filled pixel $P'_2 \equiv (240,155)$. The filling strategy is applied according \eqref{filled} and the background-target concentrations listed in Table \ref{Tab:tabella_filling_strategy}.}
	\label{fig:Reflectance_pixel_filled}
\end{figure}

Particularly, we consider multiple configurations of background-target concentrations as shown by the values of $\balpha_n$ in Table \ref{Tab:tabella_filling_strategy}: the concentrations of interest are set for background values from 0.6 to 0.1 and correspond to a cumulative target abundance in between 0.4 and 0.9, respectively. Figure \ref{fig:Reflectance_pixel_filled} shows the spectral reflectance of the filled pixel $P'_2$ for the different configurations of background-target concentrations.

Therefore, we process the filled pixel $P'_2$ in all the considered configurations for both the heuristic and constrained detectors using the spectral libraries $\bar{\bT}$ and $\bT$. In all these analyses, we verified that the output of detector is above its reference threshold, which means that the multiple sub-pixel target V$_{2}$ is correctly detected. Furthermore, Table \ref{Tab:tabella_filling_strategy} shows the estimates of the concentrations of each component of the target V$_{2}$, i.e., $\widehat{\alpha}_{2c}$ and $\widehat{\alpha}_{2b}$. 
From the table, we observe estimates for both the heuristic and constrained approaches very close to the true abundance values, especially for low background values. It is also important to remember that when using the three signatures' spectral matrix, also the abundance estimate of the third endmember, i.e., $\bt_{3}$, is provided. In this case,  the values of $\widehat{\alpha}_3$ are zero or negligible, as expected.  Furthermore, the greater spectral library seems to not influence the estimation performance, at least for the analysed cases, obtaining results comparable to those of the two signatures' spectral matrix.

\begin{table*}[!t]
	\begin{center}
		\captionof{table}{Abundances estimation over filled pixel $P_2'$ for different configurations of background-target concentrations defined according to \eqref{filled}. The $\bar{\boldsymbol{T}}$ and  $\boldsymbol{T}$ spectral matrices are applied.} 
		\begin{tabular}{c || c c c}
			$\balpha_{n}=[\alpha_{2c},\alpha_{2b}]^T$				\quad	& \quad 	&\quad 	$(\widehat{\alpha}_{2c},\widehat{\alpha}_{2b})$  	  		\quad		& $(\widehat{\alpha}_{2c},\widehat{\alpha}_{2b},\widehat{\alpha}_{3})$ \\
			\hline \hline
			\multirow{2}{*}{$\balpha_{1}=[0.35, 0.05]^T$} 					\quad	& \quad	Heuristic	&\quad 	$(0.237, 0.123)$  	\quad   	& $(0.289, 0.038, 0.022)$  \\
			\cline{2-4}
																					& \quad Constrained	&\quad 	$(0.238, 0.114)$\ 	\quad   	& $(0.292, 0.065, 0.000)$  \\
			\hline
			\multirow{2}{*}{$\balpha_{2}=[0.40, 0.10]^T$} 					\quad	& \quad	Heuristic	&\quad 	$(0.328, 0.132)$  	\quad   	& $(0.263, 0.148, 0.039)$  \\
			\cline{2-4}
																					& \quad Constrained	&\quad 	$(0.279, 0.181)$\ 	\quad   	& $(0.334, 0.130, 0.000)$  \\
			\hline
			\multirow{2}{*}{$\balpha_{3}=[0.45, 0.15]^T$} 					\quad	& \quad	Heuristic	&\quad 	$(0.367, 0.203)$  	\quad   	& $(0.269, 0.196, 0.055)$  \\
			\cline{2-4}
																					& \quad Constrained	&\quad 	$(0.298, 0.270)$\ 	\quad   	& $(0.379, 0.192, 0.000)$  \\

			\hline
			\multirow{2}{*}{$\balpha_{4}=[0.50, 0.20]^T$} 					\quad 	& \quad	Heuristic	&\quad 	$(0.368, 0.312)$  	\quad   	& $(0.398, 0.285, 0.000)$  \\
			\cline{2-4}
																					& \quad Constrained	&\quad 	$(0.359, 0.317)$\ 	\quad   	& $(0.474, 0.204, 0.000)$  \\
			\hline
			\multirow{2}{*}{$\balpha_{5}=[0.55, 0.25]^T$} 					\quad 	& \quad	Heuristic	&\quad 	$(0.465, 0.315)$  	\quad   	& $(0.495, 0.267, 0.018)$  \\
			\cline{2-4}
																					& \quad Constrained	&\quad 	$(0.447, 0.336)$\ 	\quad   	& $(0.561, 0.215, 0.007)$  \\
			\hline
			\multirow{2}{*}{$\balpha_{6}=[0.60, 0.30]^T$} 					\quad 	& \quad	Heuristic	&\quad 	$(0.536, 0.354)$  	\quad   	& $(0.580, 0.272, 0.038)$  \\
			\cline{2-4}
																					& \quad Constrained	&\quad 	$(0.536, 0.355)$\ 	\quad   	& $(0.640, 0.223, 0.025)$  \\
		\end{tabular}
		\label{Tab:tabella_filling_strategy}
	\end{center}
\end{table*}

\section{Conclusions}\label{conclusions}
In this paper, we have addressed the detection of sub-pixel targets in hyperspectral images. 
As first step, we have introduced a generalization of the so-called replacement model that includes multiple spectral signatures
with a constraint on the sum of their abundances. It is important to underline that
such a model is different from the approximate additive model that is used by most of conventional algorithms.
Then, under this generalized model, we have formulated the endmember detection problem as
a binary hypothesis test and applied GLRT-like design criteria. Specifically,
due to the intractable mathematics, we have suitably modified the maximum likelihood approach
to come up with cyclic estimation procedures. The first procedure heuristically
incorporates the constraint on the abundances whereas the second approach
exploits the Lagrange multiplier method.
Finally, we have assessed their detection and estimation performance
over synthetic and real-recorded data. As term of comparison, we have considered
the so-called ACUTE detector proposed in \cite{8964590} that, however, has been
devised under the assumption of only one spectral signature in the pixel under test.
The numerical examples have highlighted the effectiveness of both the proposed
approaches with the detector based on the Lagrange multipliers overcoming the other
counterparts.

Future research tracks might encompass the design of detectors that assume the second-order
model for the endmember signatures or are fed by contiguous pixels.


\begin{appendices}
\section{Proof of \eqref{eq:afterAppendixA}}
\label{appA}
Let us consider the matrix argument of the determinant in \eqref{eq:beforeAppendixA} and observe that
it can be written as
\begin{align}
&\left(\frac{1}{A}\bx-\widehat{\bmu}\right)
\left(\frac{1}{A}\bx-\widehat{\bmu}\right)^{T}
+\sum\limits_{k=1}^{K} \left(\bz_k - \widehat{\bmu}\right)\left(\bz_k - \widehat{\bmu}\right)^{T}
\\
&=\,
\frac{1}{A^2}\bx\bx^T+ \left(K+1\right)\widehat{\bmu} \widehat{\bmu}^T-\left(\frac{1}{A}\bx+\tilde{\bz}\right)\widehat{\bmu}^T
\nonumber
\\
&-\widehat{\bmu}\left(\frac{1}{A}\bx+\tilde{\bz}\right)^T+\bZ\bZ^T
\\
&=\,
\frac{1}{A^2}\bx\bx^T-\left(K+1\right)\widehat{\bmu} \widehat{\bmu}^T+\bZ\bZ^T
\\
&=\,
\frac{1}{A^2}\bx\bx^T-\frac{1}{\left(K+1\right)}\left(\frac{1}{A}\bx+\tilde{\bz}\right)\left(\frac{1}{A}\bx+\tilde{\bz}\right)^T\!+\bZ\bZ^T
\\
&=\,
\frac{1}{A^2}\,\bx\,\bx^T+\bZ\bZ^T-\frac{1}{K+1}\frac{1}{A^2}\,\bx\,\bx^T-\frac{1}{K+1}\,\tilde{\bz}\,\tilde{\bz}^T
\nonumber
\\
&-\frac{1}{K+1}\frac{1}{A}\,\bx\,\tilde{\bz}^T-\frac{1}{K+1}\frac{1}{A}\,\tilde{\bz}\,\bx^T
\\
&=\,
\left(\bZ\bZ^T-\frac{1}{K+1}\tilde{\bz}\,\tilde{\bz}^T\right)+\frac{K}{(K+1)}\frac{1}{A^2}\bx\,\bx^T
\nonumber
\\
&-\frac{K}{K+1}\frac{1}{KA}\,\bx\,\tilde{\bz}^T-\frac{K}{K+1}\frac{1}{KA}\,\tilde{\bz}\,\bx^T
\pm\frac{1}{K(K+1)}\tilde{\bz}\,\tilde{\bz}^T\\
&=\,
\bS_1+\frac{K}{K+1}\left(\frac{1}{A}\bx-\frac{1}{K}\tilde{\bz}\right)\left(\frac{1}{A}\bx-\frac{1}{K}\tilde{\bz}\right)^T,
\end{align}
where 
\begin{equation}
\bS_1=\left(\bZ\bZ^T-\frac{1}{K+1}\tilde{\bz}\,\tilde{\bz}^T\right)-\frac{1}{K(K+1)}\tilde{\bz}\,\tilde{\bz}^T.
\end{equation}
Exploiting the fact that $\det(\bI+\bA\bB)=\det(\bI+\bB\bA)$, $\bA\in\C^{N\times M}$ and $\bB\in\C^{M\times N}$, we can write
\begin{align}
&\det\left[
\bS_1+\frac{K}{K+1}\left(\frac{1}{A}\bx-\frac{1}{K}\tilde{\bz}\right)\left(\frac{1}{A}\bx-\frac{1}{K}\tilde{\bz}\right)^T
\right]
\nonumber
\\
&=\det\left(\bS_1\right)\nonumber
\\
&\times
\det\left[
\bI+\frac{K}{K+1}\bS_1^{-1/2}\left(\frac{1}{A}\bx-\frac{1}{K}\tilde{\bz}\right)\left(\frac{1}{A}\bx-\frac{1}{K}\tilde{\bz}\right)^T
\bS_1^{-1/2}\right]
\nonumber
\\
&=\det\left(\bS_1\right)
\left[
1+\frac{K}{K+1}\left(\frac{1}{A}\bx-\frac{1}{K}\tilde{\bz}\right)^T
\bS_1^{-1}
\left(\frac{1}{A}\bx-\frac{1}{K}\tilde{\bz}\right)\right]
\end{align}
and the proof is complete.

\end{appendices}


\bibliographystyle{IEEEtran}
\bibliography{biblio}

\end{document}